\documentstyle[12pt]{article}
\input psfig.sty
\def\Re{{\cal R \mskip-4mu \lower.1ex \hbox{\it e}\,}}
\def\Im{{\cal I \mskip-5mu \lower.1ex \hbox{\it m}\,}}
\def\ie{{\it i.e.}}

\def\etal{{\it et al.}}
\def\ibid{{\it ibid}.}
\def\sub#1{_{\lower.25ex\hbox{$\scriptstyle#1$}}}
\def\sul#1{_{\kern-.1em#1}}
\def\sll#1{_{\kern-.2em#1}}
\def\sbl#1{_{\kern-.1em\lower.25ex\hbox{$\scriptstyle#1$}}}
\def\ssb#1{_{\lower.25ex\hbox{$\scriptscriptstyle#1$}}}
\def\sbb#1{_{\lower.4ex\hbox{$\scriptstyle#1$}}}

\def\gev{\,{\rm GeV}}

\def\to{\rightarrow}
\def\slash{\not\!}
\def\mh{\ifmmode m\sbl H \else $m\sbl H$\fi}
\def\mch{\ifmmode m_{H^\pm} \else $m_{H^\pm}$\fi}
\def\mt{\ifmmode m_t\else $m_t$\fi}
\def\mc{\ifmmode m_c\else $m_c$\fi}
\def\mz{\ifmmode M_Z\else $M_Z$\fi}
\def\mw{\ifmmode M_W\else $M_W$\fi}
\def\mws{\ifmmode M_W^2 \else $M_W^2$\fi}
\def\mhs{\ifmmode m_H^2 \else $m_H^2$\fi}
\def\mzs{\ifmmode M_Z^2 \else $M_Z^2$\fi}
\def\mts{\ifmmode m_t^2 \else $m_t^2$\fi}
\def\mcs{\ifmmode m_c^2 \else $m_c^2$\fi}
\def\mchs{\ifmmode m_{H^\pm}^2 \else $m_{H^\pm}^2$\fi}
\def\ztwo{\ifmmode Z_2\else $Z_2$\fi}
\def\zone{\ifmmode Z_1\else $Z_1$\fi}
\def\mtwo{\ifmmode M_2\else $M_2$\fi}
\def\mone{\ifmmode M_1\else $M_1$\fi}
\def\bsg{\ifmmode B\to X_s\gamma\else $B\to X_s\gamma$\fi}
\def\bsll{\ifmmode B\to X_s\ell^+\ell^-\else $B\to X_s\ell^+\ell^-$\fi}
\def\bstt{\ifmmode B\to X_s\tau^+\tau^-\else $B\to X_s\tau^+\tau^-$\fi}
\def\shat{\ifmmode \hat s\else $\hat s$\fi}
\def\tb{\ifmmode \tan\beta \else $\tan\beta$\fi}
\def\xw{\ifmmode x\sub w\else $x\sub w$\fi}
\def\ch{\ifmmode H^\pm \else $H^\pm$\fi}
\def\lum{\ifmmode {\cal L}\else ${\cal L}$\fi}
\def\inpb{\ifmmode {\rm pb}^{-1}\else ${\rm pb}^{-1}$\fi}
\def\infb{\ifmmode {\rm fb}^{-1}\else ${\rm fb}^{-1}$\fi}
\def\epem{\ifmmode e^+e^-\else $e^+e^-$\fi}
\def\ppb{\ifmmode \bar pp\else $\bar pp$\fi}

\newskip\zatskip \zatskip=0pt plus0pt minus0pt
\def\matth{\mathsurround=0pt}

\def\gsim{\mathrel{\mathpalette\atversim>}}
\def\atversim#1#2{\lower0.7ex\vbox{\baselineskip\zatskip\lineskip\zatskip
  \lineskiplimit 0pt\ialign{$\matth#1\hfil##\hfil$\crcr#2\crcr\sim\crcr}}}

\renewcommand{\thefootnote}{\fnsymbol{footnote}}

\begin{document} \begin{titlepage}
\rightline{\vbox{\halign{&#\hfil\cr
&SLAC-PUB-95-6820\cr
&Revised\cr
&June 1995\cr}}}
\vspace{1in}
\begin{center}

{\Large\bf Tau Polarization Asymmetry in $B\to X_s\tau^+\tau^-$}
\footnote{Work supported by the Department of
Energy, Contract DE-AC03-76SF00515}
\medskip

\normalsize JoAnne L. Hewett
\\ \smallskip
Stanford Linear Accelerator Center,
Stanford University, Stanford, CA  94309\\
\smallskip

\end{center}

\begin{abstract}

Rare $B$ decays provide an opportunity to probe for new physics beyond the
Standard Model.  In this paper, we propose to measure the tau polarization
in the inclusive decay $B\to X_s\tau^+\tau^-$ and discuss how it can be used,
in conjunction with other observables, to completely determine the parameters
of the flavor-changing low-energy effective Hamiltonian.  Both the Standard
Model and several new physics scenarios are examined.  This process has a
large enough branching fraction, $\sim {\rm few}\times 10^{-7}$, such that
sufficient statistics will eventually be provided by the B-Factories currently
under construction.

\end{abstract}

\renewcommand{\thefootnote}{\arabic{footnote}} \end{titlepage}
%%%%%%%%%%%%%%%%%%%%%%%%%%%%%%%---- text

The recent first observation\cite{cleo} of the inclusive and
exclusive radiative decays $B\to X_s\gamma$ and $B\to K^*\gamma$ have
placed the study of rare $B$ decays on a new footing.  These flavor
changing neutral current (FCNC) transitions provide
fertile testing ground for the Standard Model (SM) and offer a
complementary strategy in the search for new physics by probing the indirect
effects of new particles and interactions in higher order processes.  In
particular, the probing of loop induced couplings can provide a means of
testing the detailed structure of the SM at the level of
radiative corrections where the Glashow-Iliopoulos-Maiani (GIM) cancellations
are important.  These first measurements have restricted the magnitude
of the electromagnetic penguin transition, resulting in bounds on the
value\cite{alickm} of the ratio of Cabibbo-Kobayashi-Maskawa (CKM) weak mixing
matrix elements $|V_{ts}/V_{cb}|$, as well as providing powerful constraints on
new physics\cite{jlhssi} which in some classes of models complement or surpass
the present bounds obtainable from direct collider searches.

The study of rare $B$ decays can be continued with the analysis of the higher
order process \bsll.  The experimental situation for these decays is very
promising\cite{bsllexp}, with \epem\ and hadron colliders closing in on the
observation of exclusive modes with $\ell=e$ and $\mu$ final states,
respectively.  These transitions proceed via electromagnetic and $Z$
penguin as well as $W$ box diagrams, and hence can probe different
coupling structures than the pure electromagnetic process \bsg.  Investigation
of this decay mode offers exciting possibilities as
various kinematic distributions associated
with the final state lepton pair, such as the lepton pair invariant mass
spectrum and the lepton pair forward-backward asymmetry, can also be measured
in addition to the total rate.  These distributions are essential in
separating the short distance FCNC processes from the contributing long
range physics\cite{ld}.  In particular, it has been
shown\cite{alione,alitwo,afb} that the lepton pair forward-backward asymmetry
is sizable for large values of the top-quark mass and is highly sensitive to
contributions from new physics.  Ali \etal\cite{alitwo} have proposed a
program to use these distributions,
as well as the total rate for \bsg, to determine the sign and magnitude of each
class of short distance FCNC contribution in a model independent
fashion.  Here, we propose a new observable,
the tau polarization asymmetry for the decay \bstt.  We will show that this
asymmetry also has a large value for top-quarks in the mass range
observed\cite{topm} at the Tevatron, and will be measurable with the high
statistics available at the B-Factories presently under construction.
The tau polarization asymmetry furnishes one more piece of available
information for the study of rare $B$ decays.  Together
with the remaining kinematic distributions mentioned above, the polarization
asymmetry (and the $M_{\tau\tau}$ spectrum) will then provide a complete
arsenal for a stringent test of the SM.

The transition rate for \bsll, including QCD corrections\cite{bsllqcd}, is
computed via an operator product expansion based on the
the effective Hamiltonian,
\begin{equation}
{\cal H}_{eff} = {4G_F\over\sqrt 2}V_{tb}V^*_{ts}\sum_{i=1}^{10}
C_i(\mu){\cal O}_i(\mu) \,,
\end{equation}
which is evolved from the electroweak scale down to $\mu\sim m_b$
by the Renormalization Group Equations.  Here $V_{ij}$ represents the
relevant CKM factors, and the ${\cal O}_i$ are a complete set of
renormalized dimension five and six operators involving light fields which
govern $b\to s$ transitions.  This basis (involving left-handed
fields only) consists of six four-quark operators ${\cal O}_{1-6}$, the
electro- and chromo-magnetic operators respectively denoted as ${\cal
O}_{7,8}$,
${\cal O}_9\sim e\bar s_{L\alpha}\gamma_\mu b_{L\alpha}\bar\ell\gamma^\mu\ell$,
and ${\cal O}_{10}\sim e\bar s_{L\alpha}\gamma_\mu b_{L\alpha}
\bar\ell\gamma^\mu\gamma_5\ell$.
%\begin{eqnarray}
%{\cal O}_1=\bar s_{L\alpha}\gamma_\mu b_{L\alpha}\bar c_{L\beta}\gamma^\mu
%c_{L\beta} \,,
%& {\cal O}_2 &=\bar s_{L\alpha}\gamma_\mu b_{L\beta}\bar c_{L\beta}\gamma^\mu
%c_{L\alpha} \,,\nonumber \\
%{\cal O}_3=\bar s_{L\alpha}\gamma_\mu b_{L\alpha}\sum_q \bar q_{L\beta}
%\gamma^\mu q_{L\beta} \,,
%& {\cal O}_4 &=\bar s_{L\alpha}\gamma_\mu b_{L\beta}\sum_q \bar q_{L\beta}
%\gamma^\mu q_{L\alpha} \,,\nonumber \\
%{\cal O}_5=\bar s_{L\alpha}\gamma_\mu b_{L\alpha}\sum_q \bar q_{R\beta}
%\gamma^\mu q_{R\beta} \,,
%& {\cal O}_6 &=\bar s_{L\alpha}\gamma_\mu b_{L\beta}\sum_q \bar q_{R\beta}
%\gamma^\mu q_{R\alpha} \,,\\
%{\cal O}_7={e\over 16\pi^2} m_b\bar si_{L\alpha}\sigma_{\mu\nu}b_{R\alpha}
%F^{\mu\nu} \,,
%& {\cal O}_8 & ={g\over 16\pi^2} m_b\bar s_{L\alpha}T^a_{\alpha\beta}
%\sigma_{\mu\nu}b_{R\alpha}G^{a\mu\nu} \,, \nonumber\\
%{\cal O}_9={e\over 16\pi^2}\bar s_{L\alpha}\gamma_\mu b_{L\alpha}
%\bar\ell\gamma^\mu\ell \,,
%& {\cal O}_{10} &={e\over 16\pi^2}\bar s_{L\alpha}\gamma_\mu b_{L\alpha}
%\bar\ell\gamma^\mu\gamma_5\ell \,, \nonumber
%\end{eqnarray}
%with $F^{\mu\nu}$ and $G^{a\mu\nu}$ being the electromagnetic and gluon
%interaction field strength tensors, respectively, and $e$ and $g$ are the
%corresponding coupling constants.
For \bsll, this effective Hamiltonian leads to the matrix element
(neglecting the strange quark mass)
\begin{equation}
M  =  {\sqrt 2 G_F\alpha\over\pi} V_{tb}V^*_{ts}\Bigg[ C_9^{eff}
\bar s_L\gamma_\mu b_L\bar\ell\gamma^\mu\ell+C_{10}\bar s_L\gamma_\mu
b_L\bar\ell\gamma^\mu\gamma_5\ell
  -2C_7 m_b\bar s_L i\sigma_{\mu\nu}{q^\nu\over q^2}b_R\bar\ell\gamma^\mu
\ell\Bigg] \,,
\end{equation}
where $q^2$ is the momentum transferred to the lepton pair.
The Wilson coefficients $C_i$ of the
$b\to s$ operators are evaluated perturbatively at the electroweak scale
where the matching conditions are imposed and are then evolved down to the
renormalization scale $\mu$.  $C_{7-10}(M_W)$ are given by the Inami-Lim
functions\cite{inlim}, $C_2(M_W)=-1$, and $C_{1,3-6}(M_W)=0$.  The
expressions for the QCD-renormalized coefficients $C_i(\mu)$ are given
explicitly in Refs. \cite{alitwo,bsllqcd}.  The effective coefficient of
${\cal O}_9$ is defined by $C_9^{eff}(\mu)\equiv C_9(\mu)+Y(\mu,q^2)$
where the function $Y$ contains the contributions from the one-loop matrix
element of the four-quark operators and can be found in Refs.
\cite{alitwo,bsllqcd}.  We note that $Y(\mu,q^2)$ contains both real and
imaginary contributions (the imaginary piece arises when the c-quarks in the
loop are on-shell).  The differential branching fraction for \bstt\ is then
\begin{eqnarray}
{dB(\bstt)\over d\shat} & = & B(B\to X\ell\bar\nu)
{\alpha^2\over 4\pi^2}
{|V_{tb}V^*_{ts}|^2\over |V_{cb}|^2} {(1-\shat)^2\over f(z)\kappa(z)}
\left[ 1-{4x\over\shat}\right]^{1/2} \Bigg\{ \left[ |C_9^{eff}|^2-|C_{10}|^2
\right] 6x \nonumber\\
& & \quad\quad\quad\quad
+\left[ |C_9^{eff}|^2+|C_{10}|^2 \right]
\left[ (\shat-4x)+(1+{2x\over\shat})(1+\shat)\right] \\
& & \quad\quad\quad\quad + 12C_7\Re C_9^{eff}(1+{2x\over\shat})
+{4|C_7|^2\over\shat}(1+{2x\over\shat})(2+\shat) \Bigg\} \,, \nonumber
\end{eqnarray}
with all Wilson coefficients evaluated at $\mu\sim m_b$, $\shat\equiv
q^2/m_b^2$, $x\equiv m_\tau^2/m_b^2$, $z\equiv m_c/m_b$, and $f(z)$ and
$\kappa(z)$ represent the phase space and QCD corrections\cite{cabb},
respectively, to the semi-leptonic rate.  This agrees with the literature
in the zero lepton mass limit.  The differential branching fraction is scaled
to that of the semi-leptonic decay $B\to X\ell\nu$ to remove the uncertainties
associated with the overall factor of $m_b^5$ and to reduce the ambiguities
involved with the imprecisely determined CKM factors.
It is well known that there are large uncertainties (up to $\pm 30\%$)
associated with the values of the coefficients $C_{7,9}(\mu)$ due to the
renormalization scale dependence of the QCD
corrections at leading-logarithmic order, as well as from the scale parameter
in $\alpha_s$.  However, this dependence is expected to be reduced at the
next-to-leading order.  This has recently been demonstrated by Buras and
M\" unz\cite{bsllqcd} for the case of $C_9(\mu)$, which was found to deviate
by only $\pm 8\%$ as the renormalization scale $\mu$ and the QCD scale
parameter $\Lambda_{\overline {MS}}$ were varied within their full range of
values.  The situation for $C_7(\mu)$ differs, however, as only partial
NLO calculations exist.  These partial calculations do exhibit a reduced $\mu$
dependence, and we eagerly await the completion of the NLO computations in
this case.

\bsll\ also receives large long distance contributions from the tree-level
processes $B\rightarrow K^{(*)}\psi^{(')}$ followed by $\psi^{(')}
\rightarrow\ell^+\ell^-$.  These pole contributions are incorporated
into the lepton pair invariant mass spectrum following the prescription in
Ref. \cite{ld}, where both on- and off-shell vector mesons are considered by
employing a Breit-Wigner form for the resonance propagator.  This produces an
additional contribution to $C_9^{eff}$ of the form
\begin{equation}
{-3\pi\over\alpha^2m_b^2}\sum_{V_i=\psi,\psi'}{M_{V_i}\Gamma(V_i\to
\ell^+\ell^-)\over (\shat-M^2_{V_i}/m_b^2)+i\Gamma_{V_i}M_{V_i}/m_b^2} \,.
\end{equation}
The relative sign between the short and long distance terms was once a source
of controversy, but can be explicitly determined via
the analyses presented in Ref. \cite{sign}.
The resulting differential branching fraction for \bsll, with and without the
long distance resonance contributions, is presented in Fig. 1a for both
$\ell=e$ and $\tau$,  taking $m_t=180\gev$, $m_b=4.87\gev$, and $z=0.316$.
We see that the pole contributions clearly overwhelm the branching fraction
near the $\psi$ and $\psi'$ peaks, and that there is significant interference
between the dispersive part of the resonance and the short distance
contributions.  However, suitable $\ell^+\ell^-$ invariant mass cuts can
eliminate the resonance contributions, and observations away from these peaks
cleanly separate out the short distance physics.  This divides the spectrum
into
two distinct regions\cite{alitwo}, (i) low-dilepton mass, $4x\leq\shat\leq
M_\psi^2/m_b^2-\delta$, and (ii) high-dilepton mass, $M^2_{\psi'}/m_b^2+
\delta\leq\shat\leq\shat_{max}$, where $\delta$ is to be matched to
an experimental cut.  The integrated branching fractions
(without the pole contributions) for $\ell=e, \mu, \tau$  are presented in
Table 1 for both the total and high dilepton mass regions of \shat.
We note that the branching fraction for \bstt\ is comparable
to that for $\ell=e,\mu$ in the clean \shat\ region above the $\psi'$
resonance.  The exclusive decay $B\to K\tau^+\tau^-$ has been computed via
heavy meson chiral perturbation theory by Du \etal\cite{bktt},
where the exclusive branching fraction was found to be
$\sim 50-60\%$ of the inclusive; this places $B(B\to K\tau^+\tau^-)$ in
the range $\sim 2\times 10^{-7}$.  Of course, calculations of exclusive
decay rates are inherently model dependent\cite{odon}, implying that some
degree
of uncertainty is associated with this result.  However, chiral perturbation
theory is known to be reliable at energy scales smaller than the typical
scale of chiral symmetry breaking, $\Lambda_{CSB}\simeq 4\pi f_\pi/\sqrt 2$.
In $B\to K\tau^+\tau^-$, the maximum energy of the $K$ meson in the $B$ rest
frame is $(m_B^2+m_K^2-4m^2_\tau)/2m_B\sim 1.5\gev$, which places most
of the available phase space at or comfortably below the
scale $\Lambda_{CSB}$.  We thus expect this method to
give a reasonable estimate of the exclusive rate.
\begin{table}
\centering
\begin{tabular}{|c|c|c|}\hline\hline
$\ell$ & $4x\leq\shat\leq 1$ & $0.6\leq\shat\leq 1$ \\ \hline
$e$ & $1.2\times 10^{-5}$ & $8.5\times 10^{-7}$ \\
$\mu$ & $1.0\times 10^{-5}$ & $8.5\times 10^{-7}$ \\
$\tau$ & $5.4\times 10^{-7}$ & $4.3\times 10^{-7}$ \\ \hline\hline
\end{tabular}
\caption{Integrated branching fractions for \bsll\ for the total and high
dilepton mass regions.}
\end{table}

The tau polarization asymmetry is defined as
\begin{equation}
P_\tau(\shat)\equiv{dB_{\lambda=-1}-dB_{\lambda=+1}\over
dB_{\lambda=-1}+dB_{\lambda=+1}}
\,,
\end{equation}
where $dB$ represents the differential \bstt\ branching fraction.
The spin projection operator is represented as $(1+\gamma_5\slash{\bf s})/2$,
with the normalized dot product being defined as $\hat{\bf s}\cdot\hat p
=\lambda=\pm 1$ with the $- (+)$ sign corresponding to the case where the
spin polarization is anti-parallel (parallel) to the direction of the
$\tau^-$ momentum.  This corresponds to the usual definition of a
polarization asymmetry, given in terms of couplings, \ie, $(L-R)/(L+R)$,
in the massless case.  We note that, of course (and unfortunately), the
polarization of final state massless leptons cannot be determined in a
collider environment.
For the process \bstt\ this
asymmetry is then calculated to be
\begin{equation}
P_\tau(\shat)={-2\left[1-4x/\shat\right]^{1/2}C_{10}\left[ \Re C_9^{eff}
(1+2\shat)+6C_7\right]\over D}\,,
\end{equation}
where $D$ is given by the expression in the curly brackets in Eq. (3).  The
tau polarization asymmetry is displayed as a function of $\shat=q^2/m_b^2$ in
Fig. 1b, with and without the long distance resonance contributions, and taking
$m_t=180\gev$.  We see that the asymmetry vanishes at threshold
and grows with increasing \shat.  The value of the total integrated asymmetry
(\ie, averaged over the high dilepton mass region,
$\shat\geq 0.6$) is $-0.484$.  The experimentally relevant number of events
required to measure an asymmetry $a$ at the $n\sigma$ level is
$N=n^2/Ba^2$, and is given
here by $N=n^2/(4.3\times 10^{-7})(-0.484)^2=(n^2)9.9\times 10^{6}$
for the inclusive decay.  The exclusive case of $B\to K\tau^+\tau^-$ would
then yield $N\sim (n^2)2.1\times 10^{7}$.
This result demonstrates that $P_\tau$ should be accessible, after several
years of running (even when $\tau$
identification efficiencies are taken into account), at the B-Factories under
construction.

The formalism for determining the polarization of $\tau$'s has been
extensively studied\cite{taupol} for many years (in fact, even before
the tau was discovered!).  More recently, polarization measurements of final
state $\tau$ leptons have been proposed as
a useful tool in discerning physics beyond the SM in a variety of processes.
Some examples include, determining the transverse and longitudinal $\tau$
polarization in $B,\Lambda_b\to X\tau\nu$\cite{yossi}, $\tau$ polarization
asymmetry at hadron colliders as a probe of new $Z'$ couplings\cite{zprime},
using $\tau$ polarization to enhance charged Higgs boson searches at hadron
colliders\cite{dproy}, probing neutralino mixing through $\tau$
polarization in scalar $\tau$ decay\cite{mihoko}, and in searching for
CP violation in the leptonic sector\cite{paul}.  These measurements can
take place as information on the tau's
polarization state is carried to its decay products.  In particular, the
momentum distribution of the decay products $A$ (in $\tau\to A\nu_\tau$, where
$A=e\bar\nu_e, \mu\bar\nu_\mu, n\pi, \rho, a_1,...$) has a large dependence on
the spin polarization state of the parent $\tau$ lepton.  This dependence is
sufficiently striking, such that the tau's helicity may be established from
a relatively low number of events\cite{taupol}.
The formalism developed for determining the tau's spin polarization on the Z
resonance can be applied in this case in the $\tau^+\tau^-$ invariant mass
rest frame, except that one cannot take the collinear limit $E_\tau\gg m_\tau$.
The resulting decay distributions cannot, however, be measured in this
rest frame due to the two undetected neutrinos, and one is forced to transform
to the laboratory frame to implement this procedure.  As an example of how
accurately the tau's polarization can be determined, we note that the four LEP
detectors have made separate polarization measurements\cite{leptau} in each of
the $\tau$ decay modes $e\nu\bar\nu, \mu\nu\bar\nu, \pi (K)\nu, \rho\nu$, and
$a_1\nu$ (in addition, DELPHI has used an inclusive one-prong hadronic
analysis).  These modes account for $\sim 80\%$ of all $\tau$ decays.  The tau
polarization has then been determined with an overall error of $10-15\%$ per
experiment.  If B-factory experiments can achieve similar results (and there
is no reason to believe otherwise) then they will have sufficient statistics
to measure the asymmetry $P_\tau$.

We now explore the sensitivity of $P_\tau$ to new physics.  We first
investigate the influence of a change in sign of the short distance
contributions to $C_{7-10}$ (holding the magnitudes constant).  The results are
shown in Fig. 2a, where the dashed, dash-dotted, dotted, solid, and
long-dashed curves represent the polarization asymmetry with
$C_{10}(M_W)\to - C_{10}(M_W), C_{9,10}(M_W)\to -C_{9,10}(M_W), C_{9}(M_W)\to
-C_9(M_W)$, the SM, and
$C_{7,8}(M_W)\to -C_{7,8}(M_W)$, respectively.  We see that there is large
sensitivity to any combination of sign changes in $C_{9,10}(M_W)$, but  little
variation to a sign change in the electro- and chromo-magnetic operator
coefficients.  This is due to the fact that the operators ${\cal O}_{9,10}$
dominate the decay in the high \shat\ region.  We next examine $P_\tau$ in
two-Higgs-Doublet models of type II, where a charged Higgs boson participates
in the decay via virtual exchange in the $\gamma, Z$ penguin  and
box diagrams.  The modifications to the Wilson coefficients in this model are
given in Deshpande \etal\cite{bsllch}.  The resulting tau polarization
asymmetry (with $\shat=0.7$ and $m_t=180\gev$) for various values of the
charged Higgs mass is presented in Fig. 2b as a function of $\tb\equiv
v_2/v_1$, the ratio of vacuum expectation values of the two doublets.
We see that the effect of the \ch\ is negligible for values
of the parameters which are consistent with the present constraints from
\bsg\cite{cleo,bsgch}, \ie, $\tb\gsim 1$ and $\mch\gsim 240 \gev$.
Finally, we study the effects of anomalous trilinear gauge boson couplings
in \bsll.  The dependence of the $C_i(M_W)$ on these anomalous couplings can
be found in Ref. \cite{bsllanom}.  Figure 2c displays
the deviation of $P_\tau$ (for $\shat=0.7$ and $m_t=180\gev$)
with non-vanishing values of the anomalous
magnetic dipole and electric quadrupole $WW\gamma$ coupling parameters,
$\Delta\kappa_\gamma$ and $\lambda_\gamma$, respectively, and of the
parameter $g_5^Z$ which governs the term $ig_5^Z\epsilon^{\mu\nu\lambda\rho}
(W^\dagger_\mu\partial_\lambda W_\nu-W_\nu\partial_\lambda W_\mu^\dagger)
Z_\rho$ in the $WWZ$ Lagrangian.  For the anomalous coupling parameters
considered here, \bsll\ naturally avoids the problem of
introducing cutoffs to regulate the divergent loop integrals due to
cancellations provided by the GIM mechanism\cite{bsllanom}.
As expected, we find little sensitivity to modifications in $C_{7,8}(M_W)$
from anomalous $WW\gamma$ couplings, but a large variation due to the
influence of anomalous $WWZ$ vertices in $C_{9,10}(M_W)$.

In order to ascertain how much quantitative information is obtainable on the
values of the Wilson coefficients $C_{7,9,10}$ from the various kinematic
distributions, we perform a
Monte Carlo study.  For illustration purposes, we will examine the case where
the SM situation is realized, \ie, we assume that there is no new physics
contributing to these decays.  For the process $B\to X_s\ell^+\ell^-$,
we consider the $M_{\ell^+\ell^-}$ distribution and the lepton pair
forward-backward asymmetry\cite{afbnote} for $\ell=e,\mu$, and $\tau$, as well
as the tau polarization asymmetry.  We take the lepton pair invariant mass
spectrum and divide it into 9 bins.  These bins are distributed as follows:
6 bins of equal size,  $\Delta\shat =0.05$, are
taken in the low dilepton mass region below the $J/\psi$
resonance, $0.02\leq \shat\leq 0.32$ (where we have also cut out
the region near zero due to the photon pole), and 3 bins in the high
dilepton mass region above the $\psi'$ pole, which are taken to be
$0.6\leq \shat\leq 0.7$, $0.7\leq\shat\leq 0.8$, and $0.8\leq\shat\leq 1.0$.
The number of events per bin is given by
\begin{equation}
N_{\rm bin}=\lum \int_{\shat_{\rm min}}^{\shat_{\rm max}} {d\Gamma \over
d\shat} \, \shat \,,
\end{equation}
and the average value of the asymmetries in each bin is then
\begin{equation}
\langle A\rangle_{\rm bin} ={\lum\over N_{\rm bin}}
\int_{\shat_{\rm min}}^{\shat_{\rm max}} A \, {d\Gamma \over d\shat} \,
\shat \,.
\end{equation}
We also include in our study the inclusive decay $B\to X_s\gamma$, which is
directly proportional to $|C_7(\mu)|^2$.
For this case we only consider the total rate.
Next we generate ``data'' (assuming the SM is correct) for an integrated
luminosity of $5\times 10^8\, B\bar B$ pairs; this corresponds to the expected
total luminosity after several years of running at future B-factories.  The
``data'' is then statistically fluctuated by a normalized Gaussian distributed
random number procedure.  The statistical errors are taken to be
$\delta N=\sqrt N$ and $\delta A=\sqrt{(1-A^2)/N}$.  We include statistical
errors only for the decay $B\to X_s\ell^+\ell^-$, as even for this large value
of integrated luminosity we expect the errors in each bin to be statistically
dominated.  The situation differs for $B\to X_s\gamma$, however, as
the statistical precision will far exceed the possible systematic (and
theoretical) accuracy.  Hence, in this case we assume a flat $10\%$ error in
the measurement of the branching fraction.  We then perform a three dimensional
$\chi^2$ fit to the coefficients $C_{7,9,10}(\mu)$ from the
``data'' according to the usual prescription
\begin{equation}
\chi^2_i=\sum_{\rm bins} \left( {Q_i^{\rm obs}-Q_i^{\rm SM}\over \delta Q_i}
\right)^2 \,,
\end{equation}
where $Q_i^{\rm obs, SM}, \delta Q_i$ represent the ``data'', the SM
expectations,
and the error for each observable quantity $Q_i$.  The resulting $95\%$ C.L.
allowed regions as projected onto
the $C_9(\mu)-C_{10}(\mu)$  and $C_7(\mu)-C_{10}(\mu)$
planes are presented in Fig. 3a and b, respectively.  In these figures, the
point `S' labels the SM expectations (assuming $m_t=180\gev$),
and the diamond represents the best fit value which has a total $\chi^2=24.6/
25{\rm dof}$.  We see that the determined ranges for the coefficients
encompasses their SM values.  The $95\%$ C.L. ranges for the coefficients
are found to be $C_7(\mu)=0.3208^{+0.0286}_{-0.0268}\,,
C_9(\mu)=-2.300^{+0.425}_{-0.495}$, and
$C_{10}(\mu)=4.834^{+0.478}_{-0.500}$, corresponding
to a $7.5\%\,, 20\%$ and $10\%$ determination of $C_7\,, C_9$, and $C_{10}$,
respectively.  Clearly, the values extracted for $C_9$ and $C_{10}$ are
highly correlated, but this is not the case for $C_7$ and $C_{10}$.
If we take $C_7(\mu)$ to have opposite sign, \ie,
$C_7(\mu)=-|C_7(\mu)|$, we find that the fit is quite poor with the best
fit yielding $\chi^2=539.1/25{\rm dof}$.

In conclusion, we have shown that measurement of the $\tau$ polarization
in \bstt\ is highly sensitive to new physics and hence provides a powerful
probe of the SM.  Together, measurement of the polarization asymmetry and the
remaining kinematic distributions associated with \bsll, will provide enough
information to completely determine the parameters of the FCNC effective
Hamiltonian.  We find that the values of the polarization can be precisely
determined with the large data samples that will be available at the
B-Factories presently under construction.
We eagerly await the completion of these machines!

\vskip.25in
\centerline{ACKNOWLEDGEMENTS}

The author thanks T.G. Rizzo for invaluable discussions, and the Phenomenology
Institute at the University of Wisconsin for their hospitality while this
work was completed.

\newpage

%
%%%%%%%%%%%%%%%%%%--- References
%%%%%%%%%%%%%%%%%%%%%%%%%%%%%%%%%%%%%%%%%%%%%%%%%%%%%%%
\def\MPL #1 #2 #3 {Mod.~Phys.~Lett.~{\bf#1},\ #2 (#3)}
\def\NPB #1 #2 #3 {Nucl.~Phys.~{\bf#1},\ #2 (#3)}
\def\PLB #1 #2 #3 {Phys.~Lett.~{\bf#1},\ #2 (#3)}
\def\PR #1 #2 #3 {Phys.~Rep.~{\bf#1},\ #2 (#3)}
\def\PRD #1 #2 #3 {Phys.~Rev.~{\bf#1},\ #2 (#3)}
\def\PRL #1 #2 #3 {Phys.~Rev.~Lett.~{\bf#1},\ #2 (#3)}
\def\RMP #1 #2 #3 {Rev.~Mod.~Phys.~{\bf#1},\ #2 (#3)}
\def\ZP #1 #2 #3 {Z.~Phys.~{\bf#1},\ #2 (#3)}

\begin{figure}[htbp]
\centerline{
\psfig{figure=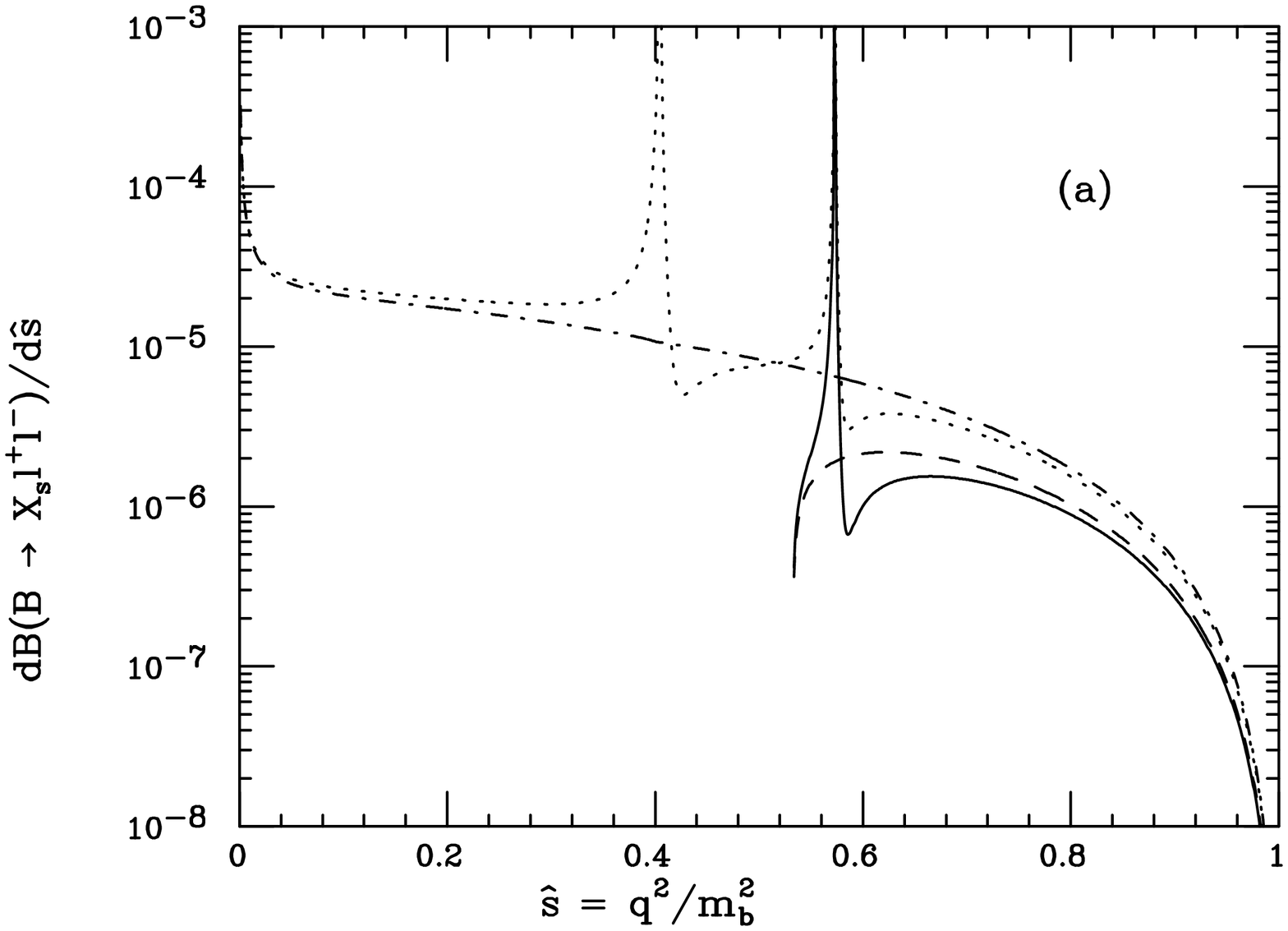,height=10cm,width=12cm,angle=90}}
\vspace*{-.75cm}
\centerline{
\psfig{figure=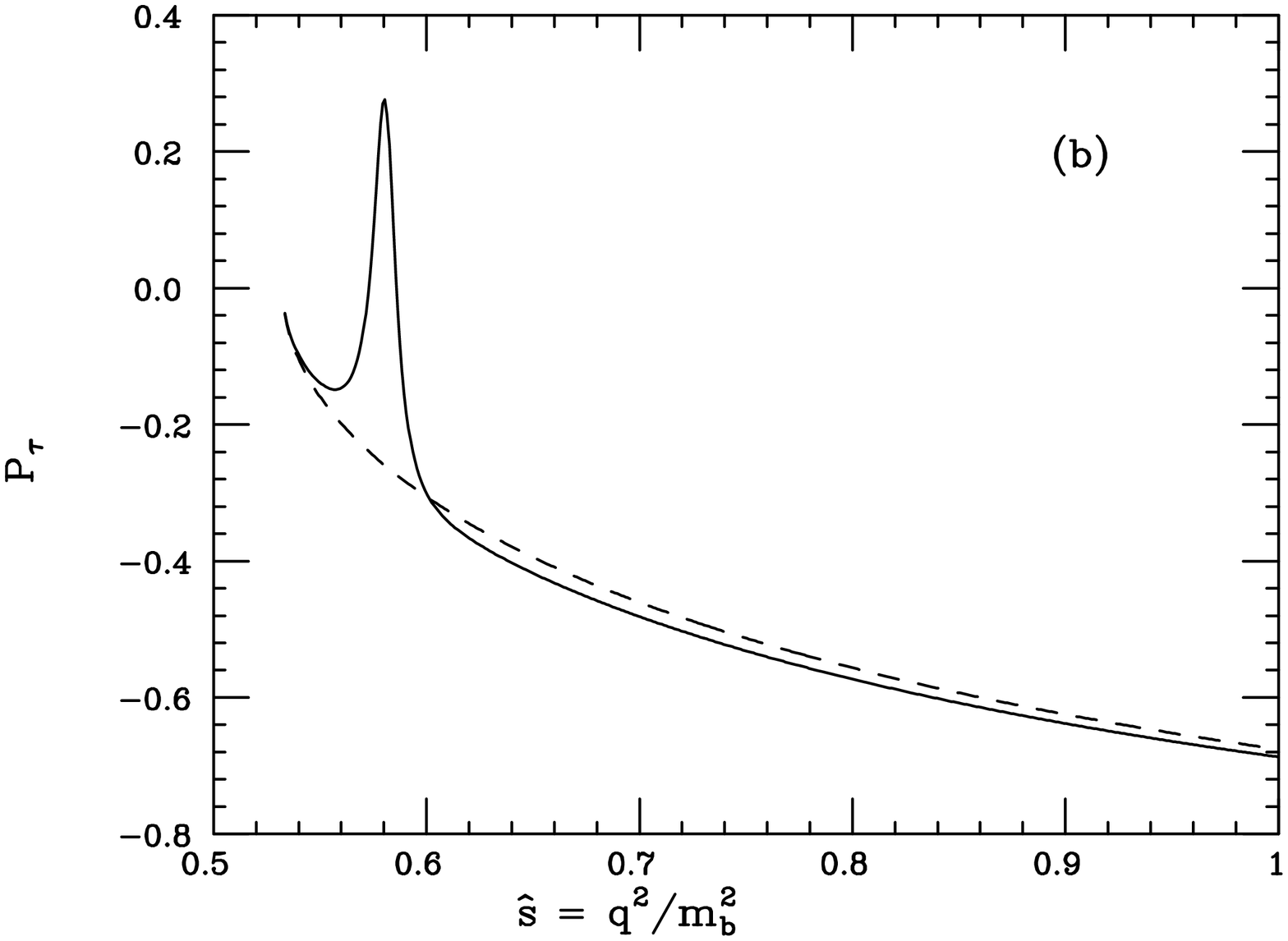,height=10cm,width=12cm,angle=90}}
\vspace*{-1cm}
\caption{\small (a) Differential branching fraction and (b) tau polarization
asymmetry as a function of \shat\ for $\ell=\tau$ (solid and dashed curves)
and $\ell=e$ (dotted and dash-dotted curves), with and without the long
distance resonance contribution.}
\end{figure}

\begin{figure}[htbp]
\centerline{
\psfig{figure=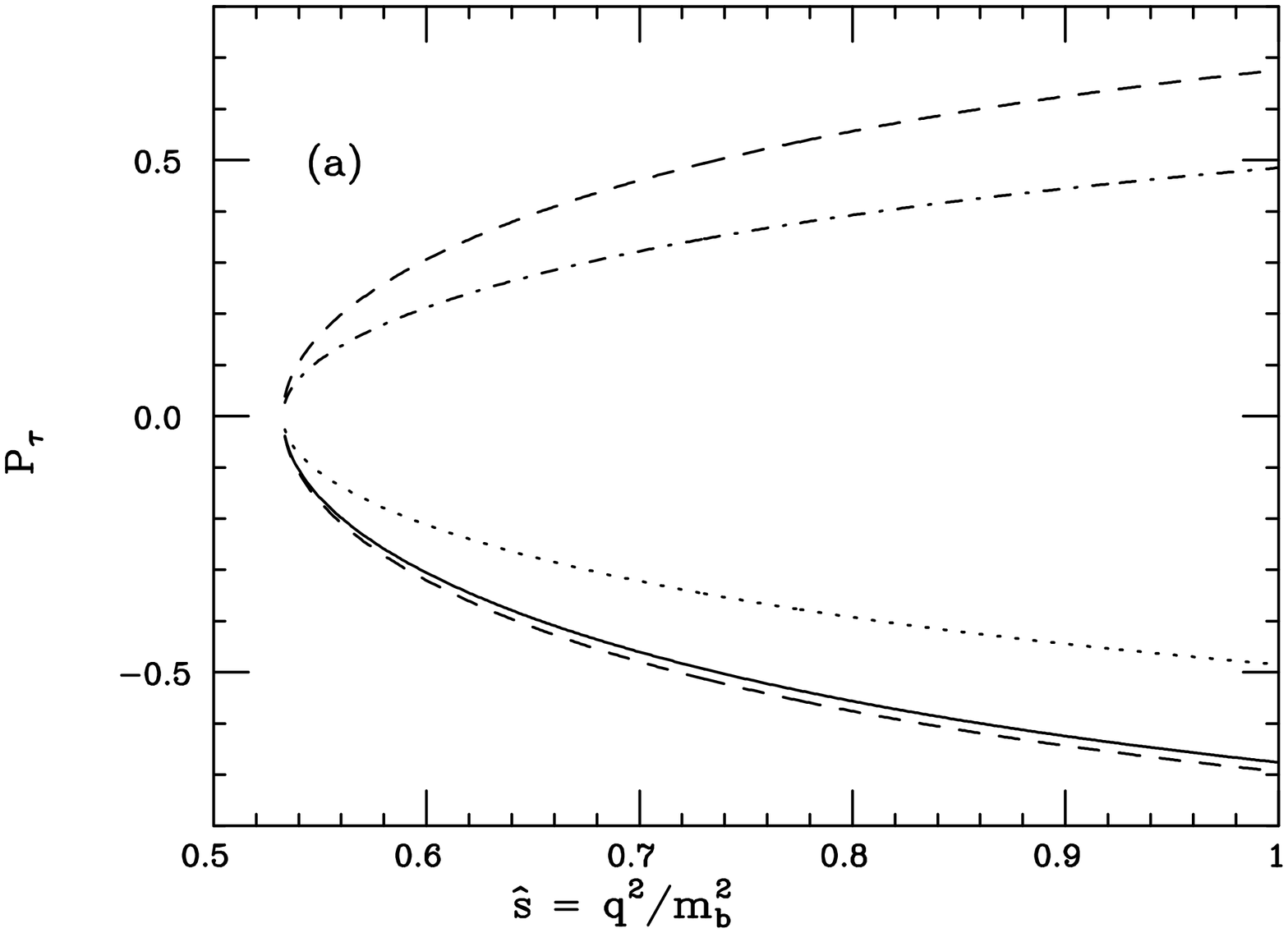,height=9.cm,width=8.8cm,angle=90}}
\vspace*{-0.75cm}
\centerline{
\psfig{figure=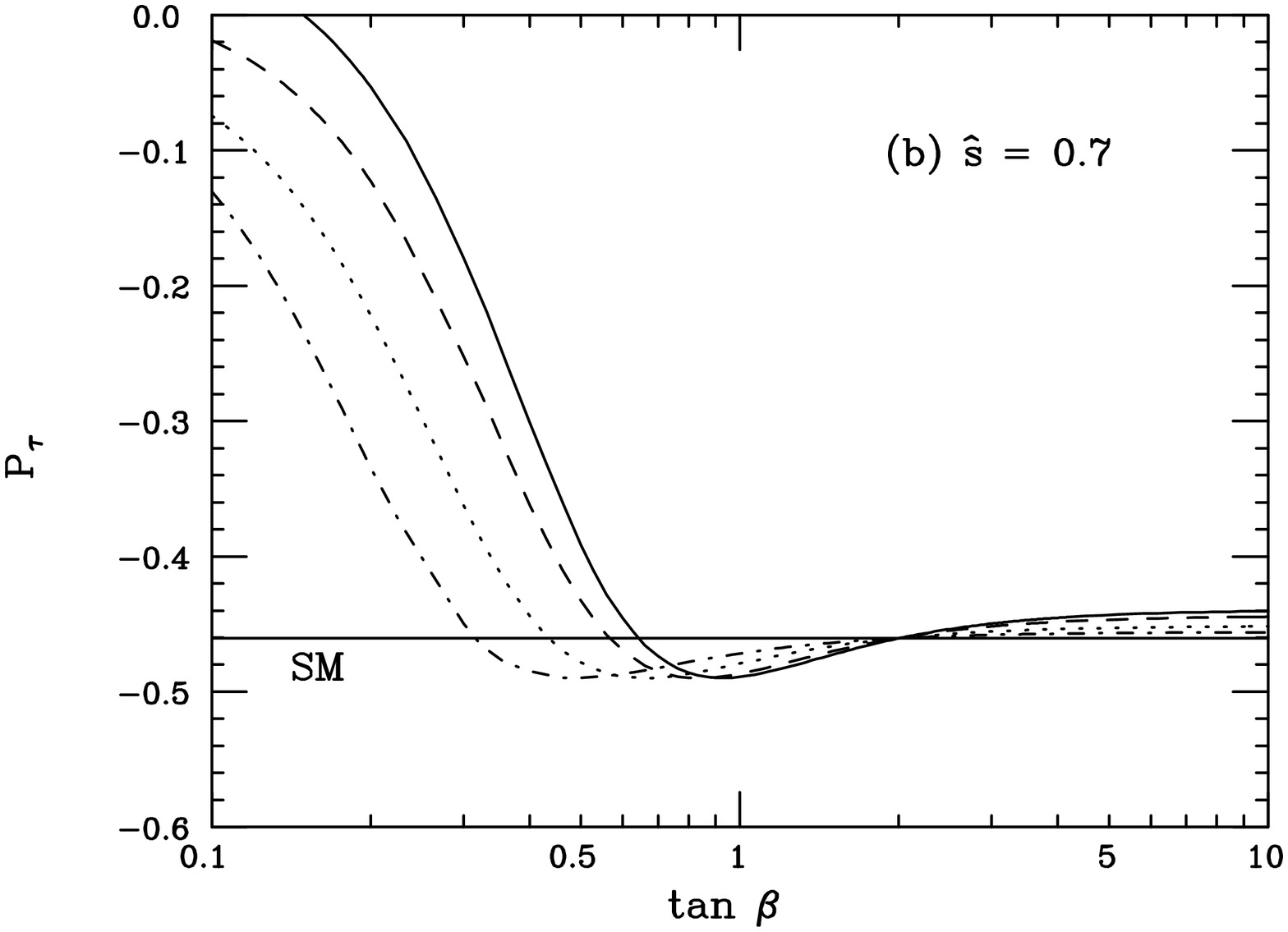,height=9.cm,width=8.8cm,angle=90}
\hspace*{-5mm}
\psfig{figure=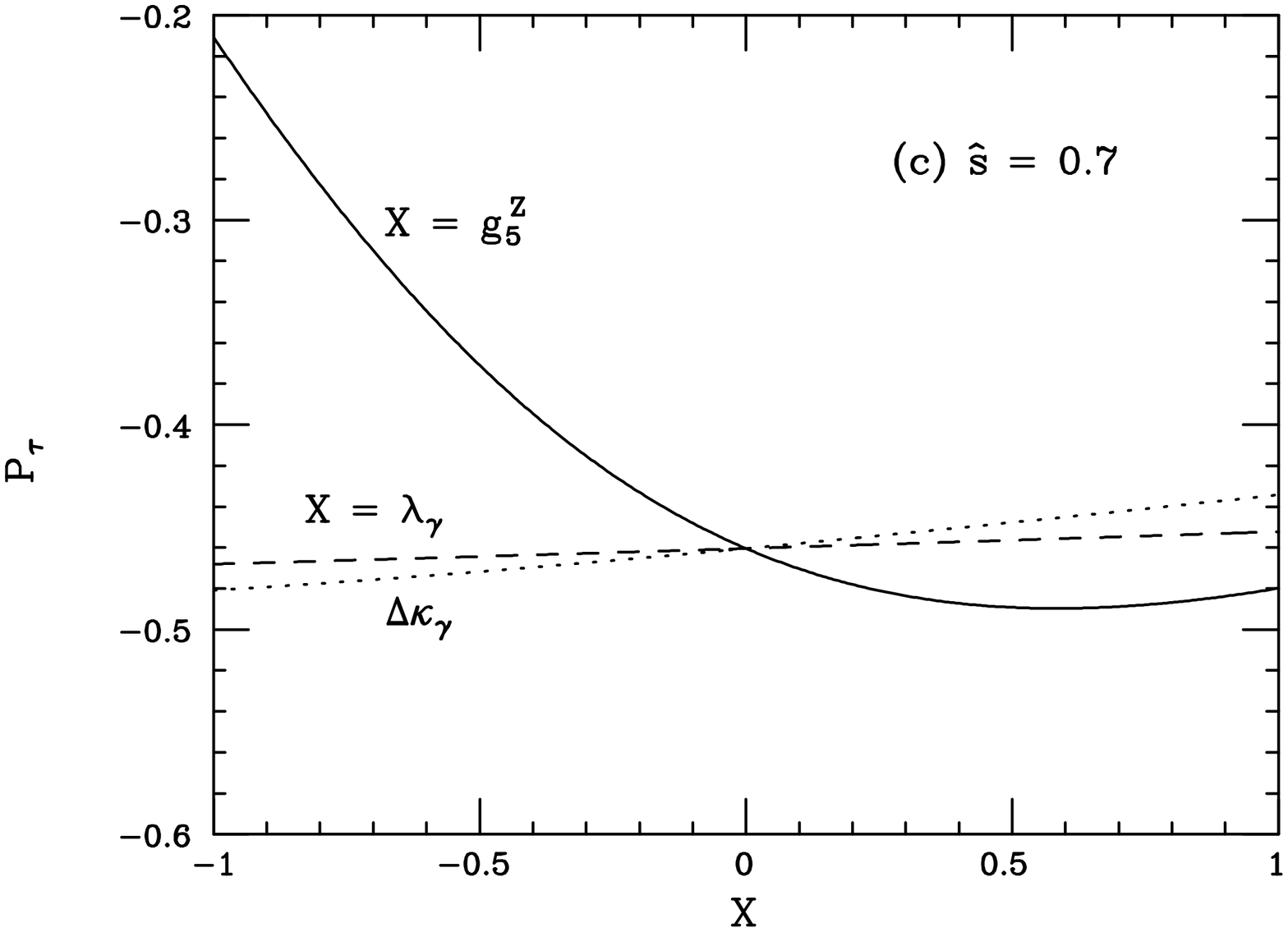,height=9.cm,width=8.8cm,angle=90}}
\vspace*{-1cm}
\caption{\small Tau polarization asymmetry (a) with changes in the sign of
the Wilson coefficients at the electroweak scale, corresponding to $C_{10}$,
$C_{9,10}$, $C_{9}$, SM $C_{7,8}$ from top to bottom; (b) in two-Higgs-doublet
models as a function of $\tan\beta$ with $\mch=50, 100, 250, 500$
corresponding to the solid, dashed, dotted, and dash-dotted curves,
respectively.  The SM value is denoted by the solid horizontal line.
(c) with anomalous couplings $WW\gamma$ and $WWZ$ couplings as described in
the text.}
\end{figure}

\begin{figure}[htbp]
\centerline{
\psfig{figure=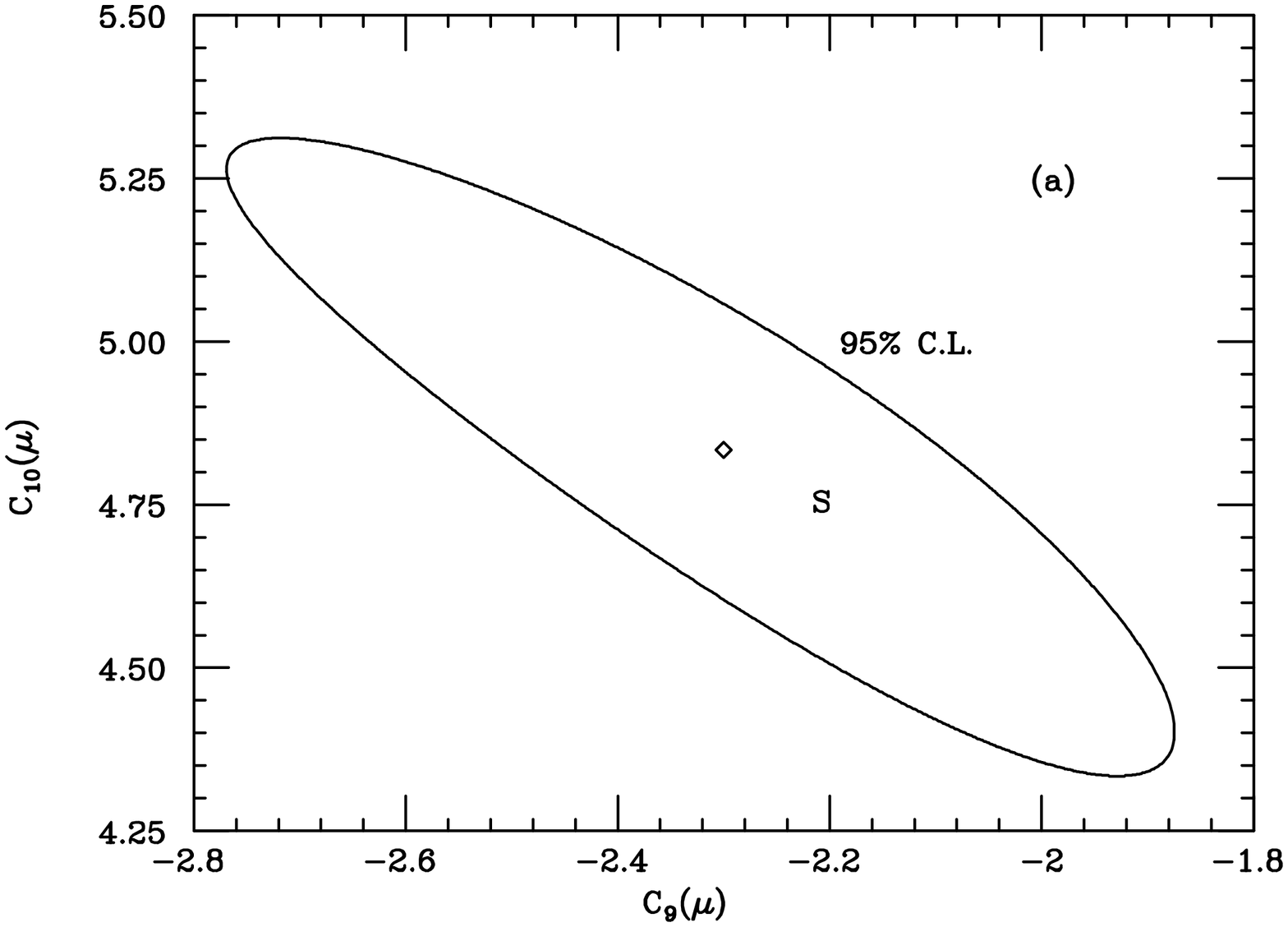,height=10cm,width=12cm,angle=-90}}
\vspace*{-.75cm}
\centerline{
\psfig{figure=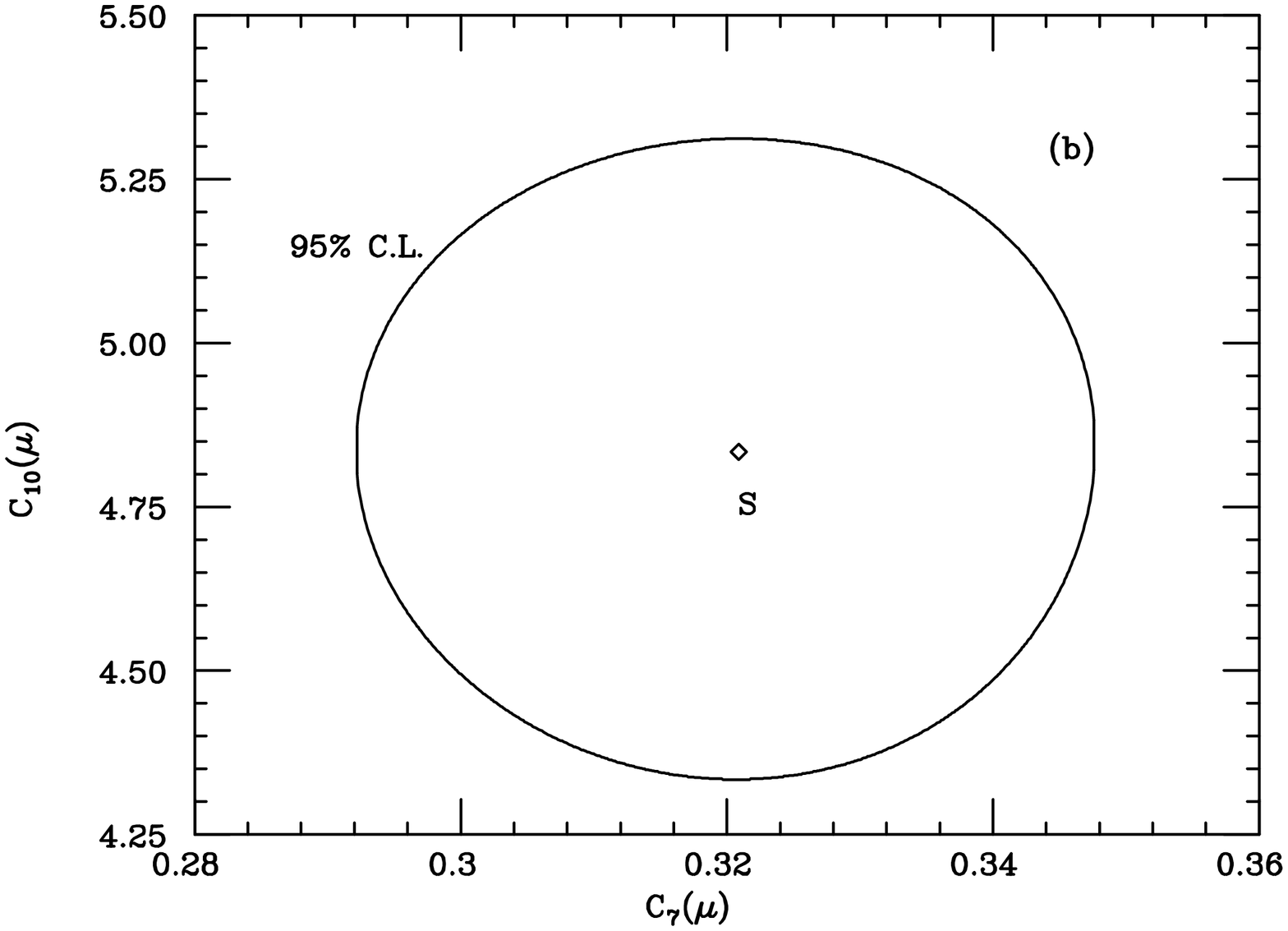,height=10cm,width=12cm,angle=-90}}
\vspace*{-1cm}
\caption{\small $95\%$ C.L. contour in the (a) $C_9 - C_{10}$, (b)
$C_7 - C_{10}$  plane from the
fit procedure described in the text.  `S' labels the SM prediction and the
diamond represents the best fit values.}
\end{figure}


\begin{thebibliography}{99}
\bibitem{afbnote}
We note that the expressions given for the lepton pair forward-backward
asymmetry in Refs. 6 and 7 do not agree; we have found agreement with that
of Ref. 6.  In the case of massive leptons, \ie, $\ell=\tau$, one must
include an additional overall factor of $[1-4x/\shat]^{1/2}$ in the
asymmetry.
\end{thebibliography}
\end{document}